\documentclass[]{jfm}

\usepackage{graphicx}
\usepackage{newtxtext}
\usepackage{newtxmath}
\usepackage{natbib}
\usepackage{hyperref}
\hypersetup{
    colorlinks = true,
    urlcolor   = blue,
    citecolor  = black,
}

\newcommand{\RomanNumeralCaps}[1]
\linenumbers

\title{\textcolor{black}{Evidence on the incompatibility of smoothed particle hydrodynamics and eddy viscosity models for large eddy simulations}}

\author{Max Okraschevski\aff{1,2}
  \corresp{\email{max.okraschevski@dlr.de}},
  Niklas Bürkle\aff{1}, Markus Wicker\aff{1},
  Rainer Koch\aff{1} \and Hans-Jörg Bauer\aff{1}}

\affiliation{\aff{1} Institute of Thermal Turbomachinery, Karlsruhe Institute of Technology, Kaiserstraße 12, 76131 Karlsruhe
\aff{2}  Present Affiliation: Institute of Engineering Thermodynamics, German
Aerospace Center, 89081 Ulm, Germany; Helmholtz
Institute Ulm for Electrochemical Energy Storage, 89081 Ulm
}

\begin{document}
\maketitle

\begin{abstract}
In this work, we will present evidence for the incompatibility of \textcolor{black}{Smoothed Particle Hydrodynamics (SPH)} methods and eddy viscosity models. Taking a coarse-graining perspective, we physically argue that SPH methods operate intrinsically as Lagrangian Large Eddy Simulations \textcolor{black}{(LES)} for turbulent flows with strongly overlapping discretization elements. However, these overlapping elements in combination with numerical errors cause a significant amount of implicit subfilter stresses (SFS). Considering a Taylor-Green flow at $\Rey=10^4$, the SFS will be shown to be relevant where turbulent fluctuations are created, explaining why turbulent flows are challenging even for \textcolor{black}{current} SPH methods. Although one might hope to mitigate the implicit SFS using eddy viscosity models, we show a degradation of the turbulent transition process, which is rooted in the non-locality of these methods.
\end{abstract}

\section{Motivation}
\label{sec:Motivation}

The Smoothed Particle Hydrodynamics (SPH) method was proposed in 1977 as a Lagrangian discretization method for fluid dynamics \citep{Lucy_1977, Gingold_1977} and matured significantly since then as detailed in several reviews \citep{Springel_2010, Price_2012, Monoghan_2012, Shadloo_2016, Ye_2019, Lind_2020, Sigalotti_2021}. Originally, it features serious numerical convergence problems due to the fact that the consistency of spatial derivative operators is strongly affected by the local particle arrangement, which only can be compensated by a drastic increase in the number of neighbor particles $N_{ngb}$ \citep{Zhu_2015}. This is especially problematic in strong shear flows and subsonic turbulence, resulting in zeroth-order errors related to excessive numerical dissipation for small $N_{ngb}$ \citep{Ellero_2010, Bauer_2012, Colagrossi_2013, Hopkins_2015}.

Pioneered by the work of \citet{Vila_1999}, \textcolor{black}{current} SPH methods, as recently compared by \citet{Eiris_2023}, mostly eliminated this convergence issue by the usage of at least one of the following two strategies 
\newline
\begin{enumerate}
    \item a consistent approximation of spatial derivatives by either a reproducing kernel (RK) or Moving-Least-Square (MLS) approach \citep{Frontiere_2017, Hopkins_2015}.
    \item \textcolor{black}{an Arbitrary-Lagrangian-Eulerian (ALE) formulation with transport velocity as noise mitigation technique \citep{Oger_2016, Antuono_2021_1}, also rigorously incorporating particle shifting \citep{Xu_2009, Lind_2012}.} \newline
\end{enumerate} 

In line with these positive developments, the confidence into the ability of \textcolor{black}{current} SPH methods to capture incompressible turbulence increases. Especially in complex multiphase flow situations, \textcolor{black}{where SPH can play to its strengths}, these methods are nowadays optimistically combined with a Large Eddy Simulation (LES) perspective \citep{Colagrossi_2021, Lai_2022, King_2023, Meringolo_2023}. In the following, we will refer to this combination of SPH methods with LES simply as SPH-LES for brevity. As already mentioned by \citet{Bicknell_1991}, such a confluence is intuitive since the SPH kernel and the LES low-pass filter, building the foundation of both methods, are mathematically congruent. Certainly, the development of rigorous combined SPH-LES theories is a very recent topic, e.g. \citet{DiMascio_2017, Antuono_2021_2, Okraschevski_2022}.

%%%%%%%%%%%%%%%%%%%%
\begin{figure}
\centering
\includegraphics[trim=0cm 0cm 0cm 0cm, clip, width=5in]{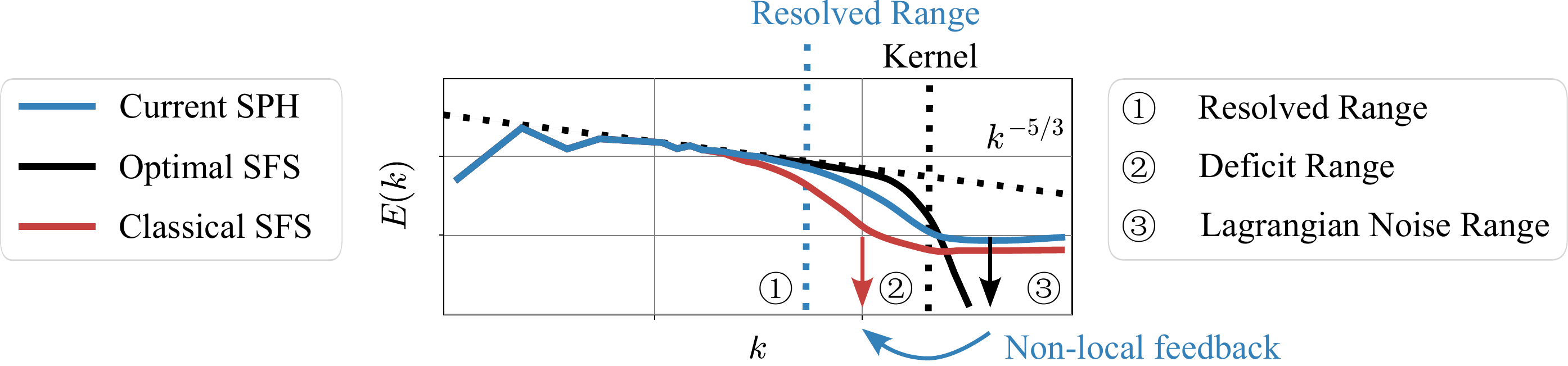}
\caption{\textcolor{black}{Typical distribution of spectral energy density obtained with SPH methods for incompressible turbulence. The properly resolved range with large eddies passes into an energy deficit range which is non-locally caused and followed by a Lagrangian noise range. From an optimal SFS model we would expect a reduction of the Lagrangian noise in favor of the deficit range. However, with incompatible classical SFS models the noise is barely reduced and the deficit range exacerbated due to non-locality.} }
\label{fig:Rev_Spectrum}
\end{figure}
%%%%%%%%%%%%%%%%%%%%

One central remaining issue is that all these \textcolor{black}{current} SPH-LES studies intuitively model one of the central LES objects, namely the subfilter stress tensor $\mathsfbi{\tau}_{SFS}$, by classical, functional eddy viscosity approaches, employing the Boussinesq hypothesis \citep{Schmitt_2007}. In \citet{Okraschevski_2022} we could argue for classical SPH, physically reinterpreting this method from a spatial coarse-graining perspective, that eddy viscosity modelling must fail due to the non-locality of the SPH method \citep{Du_2020, Vignjevic_2021, Yao_2022}. \textcolor{black}{The resulting incompatibility is illustrated in figure \ref{fig:Rev_Spectrum} and manifests in a spectral mismatch where the classical SFS model is introduced.}  Yet, one might oppose that this is just a consequence of the classical SPH approach considered in our former study. This is where the following work comes into play showing that our coarse-graining perspective generally applies to \textcolor{black}{current} SPH methods \footnote{\textcolor{black}{The aforementioned convergence issues with SPH, the countermeasures listed above and the subsequent rationale strictly apply only to Lagrangian SPH methods, inherently including particle disorder. For regular particle distributions in an Eulerian frame the convergence issues to begin with can be fully eliminated \citep{Hopkins_2015, Lind_2016} with proper numerical schemes (Appendix \ref{appA}) and, accordingly, the spectral peculiarities in figure \ref{fig:Rev_Spectrum}. However, since we believe that the Lagrangian character is still a key argument for \textcolor{black}{current} SPH methods, we will subsequently focus on this specific reference frame.}}. Hence, classical eddy viscosity modelling, \textcolor{black}{as already indicated by \citet{Rennehen_2021},} is compromising the most accurate prediction of turbulence possible and necessitates the development of completely new and specific subfilter stress models in the SPH-LES context. \textcolor{black}{This could perspectively even more accentuate the advantages of \textcolor{black}{current} SPH methods over traditional grid-based approaches in application areas like multiphase flows encompassing turbulence. }

\textcolor{black}{However, should such an improved SFS model not be available soon, we definitively \textcolor{black}{advise} to operate \textcolor{black}{current} SPH methods for aforementioned flows using no explicit SFS model. Thus, we will broadly argue in favor of the notion that \textcolor{black}{current} Lagrangian SPH methods operate intrinsically as implicit LES. 
The latter is an established concept in the grid-based community and relies on properly designed discretization errors to provide an implicit SFS contribution, e.g. \cite{Grinstein_2007, Moura_2017, Dairay_2017, Fehn_2022, Volpiani_2024}. Heuristically, it seems that these implicit SFS in SPH methods emerge from statistical physics principles \citep{Posch_1995, Ellero_2010, Borreguero_2019, Okraschevski_2021_1}, which we will confirm in more detail below using Hardy's theory \citep{Hardy_1982}. }%It is historically interesting to note that not only Hardy's theory \citep{Hardy_1982}, but also classical SPH \citep{Monaghan_1983} and the first implicit LES approaches in the grid-based community \citep{Grinstein_2007} were all established in the context of shock flows.}

\section{Novelty \& Implications}
\label{sec:Novelty}

In our former works \citep{Okraschevski_2021_1, Okraschevski_2021_2} we laid the theoretical foundations to be summarized in Sec. \ref{sec:Theory} with Hardy's theory \citep{Hardy_1982} at its core. Based on it, we could demonstrate that SPH-LES with classical SPH seems fundamentally incompatible with explicit viscosity models due to the non-local characteristic of the discretization \citep{Okraschevski_2022, Okraschevski_2024}. Here, we report evidence for the first time that this incompatibility also holds for SPH-LES with \textcolor{black}{current} SPH methods, which are not plagued by the classical SPH problems as described in Sec. \ref{sec:Motivation}. As a consequence, we believe that classical eddy viscosity models for the subfilter stress tensor $\mathsfbi{\tau}_{SFS}$ are detrimental in SPH-LES of incompressible turbulence and that novel models matching the discretization characteristics must be developed. \textcolor{black}{This insight will particularly improve the predictive power of \textcolor{black}{current} SPH methods in complex multiphase flows, when incompressible turbulence is an inevitable aspect of the considered flow.}

\section{Coarse-graining perspective on SPH methods}
\label{sec:Theory}

\begin{figure}
\centering
\includegraphics[trim=0cm 0cm 0cm 0cm, clip, width=5in]{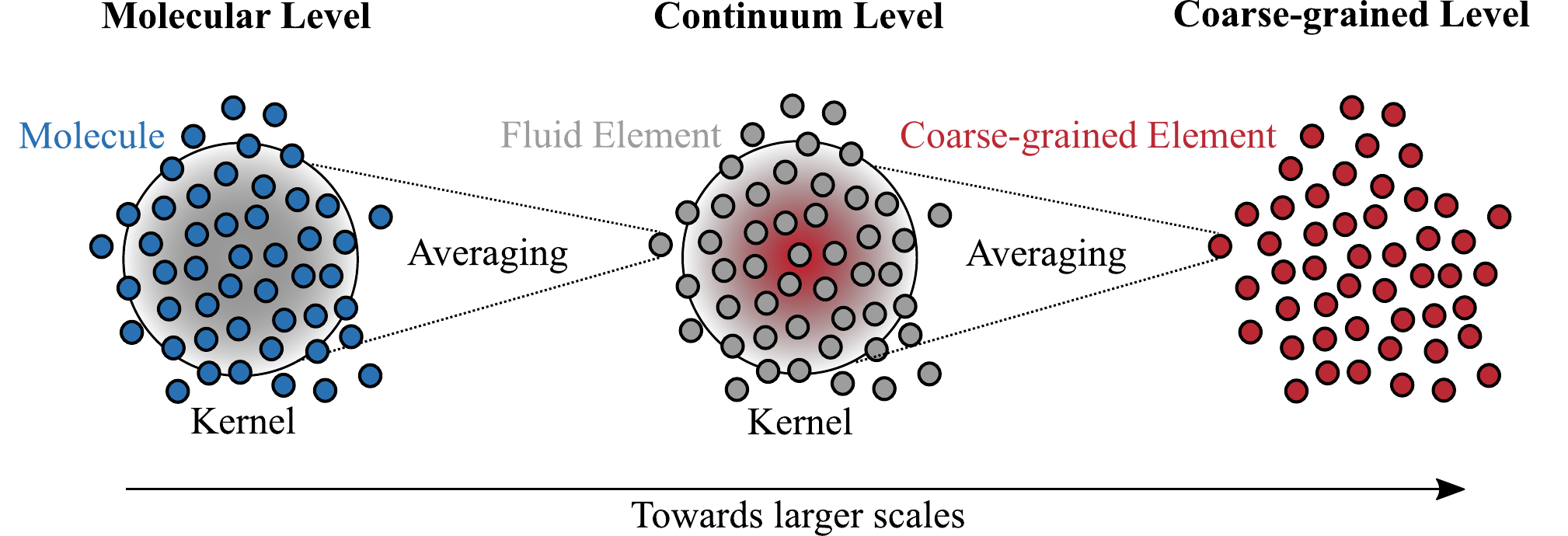}
\caption{Illustration of spatial coarse-graining emerging from the generalization of Hardy's theory \citep{Hardy_1982, Okraschevski_2021_2}. Adapted from \citet{Okraschevski_2022}.  }
\label{fig:00_CoarseGraining}
\end{figure}

Although \textcolor{black}{current} SPH methods might not suffer from the same convergence issue as their original ancestor, at their heart they still employ quasi-Lagrangian \footnote{This term is usually used within ALE frameworks \citep{Vogelsberger_2012, Oger_2016, Antuono_2021_2}. Here, we also include pseudo-Lagrangian \citep{Vogelsberger_2012} particles, also called purely Lagrangian particles in other works \citep{Oger_2016,Antuono_2021_2}, into the definition.} particles. These are connected by a spherical, positive, and monotonously decaying kernel $W: \mathbb{R}^3 \to \mathbb{R}$ with compact support $V_{\boldsymbol{x}}$, being centered at the quasi-Lagrangian particles in $\boldsymbol{x} \in \mathbb{R}^3$. Hence, even \textcolor{black}{current} SPH methods intrinsically contain two resolution scales, namely the mean particle distance $\Delta l$ and the larger kernel diameter $D_K$. We termed this property \emph{particle duality} in our former work \citep{Okraschevski_2022}. This is a characteristic peculiarity compared to grid-based discretization techniques and results in strongly overlapping discretization elements, i.e. a non-local discretization. Despite the fact that numerical convergence in \textcolor{black}{current} SPH methods can be reached using a constant ratio of $D_K/\Delta l=\textit{O}(1)$ \citep{Vila_1999, Hopkins_2015} resulting in a fixed number of neighbors $N_{ngb}$ inside the kernel, one still might wonder which flow scales can be effectively resolved. Taking a conservative point of view, it must be expected that flow scales can be captured maximally up to the kernel diameter $D_K$. By means of such a rational, one implicitly interprets SPH methods from a spatial coarse-graining perspective at the effective scale $D_K$. Such a coarse-graining perspective is not only a convenient footing in the following, but also the physical foundation of the LES community \citep{Eyink_2018} and even more so a general perspective employed by fluid dynamicists \citep{Irving_1950, Okraschevski_2021_2, Eyink_2024}. In the LES community the coarse-graining is also synonymously called \emph{the filtering approach} \citep{Germano_1992}. Although the different terminologies describe equivalent mathematical operations, we believe that the term coarse-graining raises the awareness for a geometric interpretation in terms of a hierarchical clustering of Lagrangian particles (figure \ref{fig:00_CoarseGraining}).
%%%%%%%%%%%%%%%%%%%%
%%%%%%%%%%%%%%%%%%%%
The aforementioned can be vividly unravelled by generalizing the theory of Hardy from statistical physics \citep{Hardy_1982, Okraschevski_2021_2} and highlights the conceptual similarity to the Lagrangian discretization techniques of interest. Hence, we anticipate that SPH methods aim at solving an effective field equation and intrinsically operate as Lagrangian Large Eddy Simulations. By defining the Lagrangian derivative as 

\begin{equation}
    \frac{\mathrm{d}}{\mathrm{d}t}:= \partial_t + \tilde{\boldsymbol{v}} \cdot \nabla_{\boldsymbol{x}}
\end{equation}

and the spatial coarse-graining of a scalar field $f: \mathbb{R}^3 \times \mathbb{R}^+_0 \to \mathbb{R}$ over $V_x$ as 

\begin{equation}
    \overline{f} (\boldsymbol{x}, t) = \int\displaylimits_{V_{\boldsymbol{x}}} f(\boldsymbol{y}, t) W(\boldsymbol{x} - \boldsymbol{y})~\mathrm{d}\boldsymbol{y}~,
    \label{eq:Average}
\end{equation}

we will subsequently consider barotropic flows in a Lagrangian reference frame  

\begin{equation}
    \frac{\mathrm{d} \overline {\rho}} {\mathrm{dt}}(\boldsymbol{x}, t) = -\overline{\rho} (\boldsymbol{x}, t) \nabla_{\boldsymbol{x}} \cdot \tilde{\boldsymbol{v}}(\boldsymbol{x}, t)
    \label{eq:Mass}
\end{equation}

\begin{equation}
    \overline{\rho} (\boldsymbol{x}, t) \frac{\mathrm{d} \tilde{\boldsymbol{v}}} {\mathrm{dt}}(\boldsymbol{x}, t)  = - \nabla_{\boldsymbol{x}} \overline{p}(\boldsymbol{x}, t) + div_{\boldsymbol{x}} \left[\tilde{\mathsfbi{\tau}}_{visc} + \mathsfbi{\tau}_{SFS} \right](\boldsymbol{x}, t)
    \label{eq:Momentum}
\end{equation}

\begin{equation}
    \overline{p}(\boldsymbol{x}, t) = \overline{p}_{ref} + c_s^2 ( \overline{\rho}(\boldsymbol{x}, t) - \overline{\rho}_{ref} ) .
    \label{eq:EOS}
\end{equation}
\newline
Moreover, we will assume for Eqs. (\ref{eq:Mass}), (\ref{eq:Momentum}), (\ref{eq:EOS}) a weakly-compressible, low Mach number flow ($Ma < 0.3$). Hence, bulk viscous stresses are neglected. The fields $\overline{\rho}$ and $\overline{p}$ denote the coarse-grained density and pressure, whereas $\tilde{\boldsymbol{v}} = \overline{\rho\boldsymbol{v}}/\overline{\rho}$ is the density-weighted coarse-grained velocity as proposed by \citet{Reynolds_1895}, nowadays called Favre averaged velocity \citep{Bilger_1975}.  For a Newtonian fluid, then, the \textcolor{black}{viscous} stress tensor reads 
\begin{equation}
    \tilde{\mathsfbi{\tau}}_{visc}=\eta(\mathsfbi{J}_{\tilde{\boldsymbol{v}}} +\mathsfbi{J}_{\tilde{\boldsymbol{v}}}^{\mathrm{T}}  - \frac{2}{3}\nabla_{\boldsymbol{x}}\cdot \tilde{\boldsymbol{v}})
    \label{eq:StressTensor}
\end{equation}
with $\mathsfbi{J}_{\tilde{\boldsymbol{v}}}$ as Jacobian of $\tilde{\boldsymbol{v}}$. The dynamic viscosity $\eta$, the reference density $\overline{\rho}_{ref}$ and pressure $\overline{p}_{ref}$, as well as the speed of sound $c_s$ are dealt as constant parameters to be specified.

\begin{figure}
\centering
\includegraphics[trim=0cm 0cm 0cm 0cm, clip, width=3in]{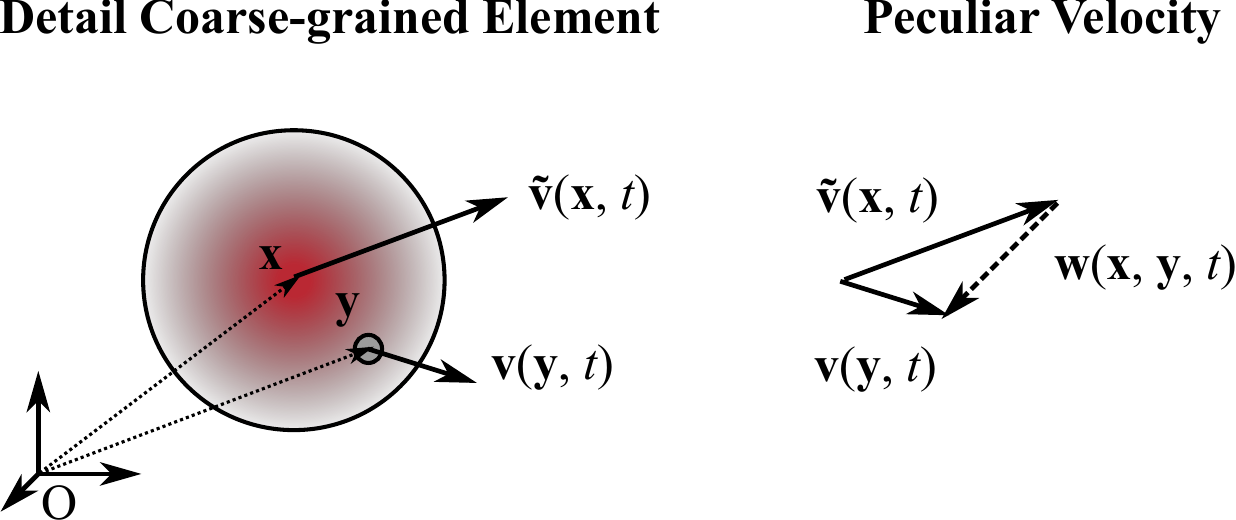}
\caption{Visualization of the velocity decomposition in Equation (\ref{eq:PeculiarVelocity}). \newline Adapted from \citet{Okraschevski_2024}.  }
\label{fig:VelocityDecomposition}
\end{figure}

The most important object emerging from the spatial coarse-graining at the arbitrary scale $D_K$ is the subfilter stress tensor $\mathsfbi{\tau}_{SFS}$. It can be formally defined as \citep{Vreman_1994, Okraschevski_2021_2}
\begin{equation}
    \mathsfbi{\tau}_{SFS} (\boldsymbol{x}, t) :=- \int\displaylimits_{V_x} \rho(\boldsymbol{y}, t) \boldsymbol{w}(\boldsymbol{x}, \boldsymbol{y}, t)\boldsymbol{w}^\mathrm{T}(\boldsymbol{x}, \boldsymbol{y}, t)  W(\boldsymbol{x} - \boldsymbol{y}) ~ \mathrm{d}\boldsymbol{y} =- \overline{\rho\boldsymbol{w}\boldsymbol{w}^\mathrm{T}}(\boldsymbol{x}, t)
    \label{eq:SubfilterStress}
\end{equation}
with $\boldsymbol{w}$ denoting the peculiar velocity and $\boldsymbol{w}^\mathrm{T}$ its transpose. The peculiar velocity is a relative velocity connecting coarse-grained velocities $\tilde{\boldsymbol{v}}(\boldsymbol{x}, t)$ with associated continuum fluid element velocities $\boldsymbol{v}(\boldsymbol{y}, t)$ (figure \ref{fig:00_CoarseGraining}). \textcolor{black}{By} convention, we label and distinguish the spatial coordinates of fluid elements by $\boldsymbol{y} \in \mathbb{R}^3$ and of coarse-grained elements by the spatial coordinate $\boldsymbol{x} \in \mathbb{R}^3$. Consequently, the peculiar velocity is a non-local quantity by definition and emerges from 
\begin{equation}
   \boldsymbol{w}(\boldsymbol{x}, \boldsymbol{y}, t) = \boldsymbol{v}(\boldsymbol{y}, t) - \tilde{\boldsymbol{v}}(\boldsymbol{x}, t) ~.
    \label{eq:PeculiarVelocity}
\end{equation}
The velocity decomposition is illustrated in figure \ref{fig:VelocityDecomposition}. We argue that $\boldsymbol{w}$ is the physically appropriate fluctuating velocity field in the coarse-graining framework. It satisfies $\overline{\rho \boldsymbol{w}}=0$ by construction \citep{Okraschevski_2021_2} and does not require the introduction of generalized central moments to identify the averaging invariance of the turbulent equations \citep{Germano_1992}. Since the velocities $\boldsymbol{v}(\boldsymbol{y}, t)$ of the fluid elements are unknown at the coarse-grained level, we practically face the well-known closure problem for the SFS tensor $\mathsfbi{\tau}_{SFS}$ in equation (\ref{eq:SubfilterStress}). 

%%%%%%%%%%%%%%%%%%%%
%%%%%%%%%%%%%%%%%%%%

For turbulent flows at finite resolution, the closure problem is \textcolor{black}{often} resolved by explicit modelling of $\mathsfbi{\tau}_{SFS}$ with functional eddy viscosity approaches \citep{Silvis_2017, Moser_2021} \textcolor{black}{assuming dominant physical SFS error over discretization error. However, as long as the filter width and resolution width are similar, it is well-known in the grid-based LES community that these errors are at least in the same order of magnitude or, \textcolor{black}{in fact, that the latter exceeds the former} \citep{Ghosal_1996, Dairay_2017}. This insight motivated the development of the nowadays well established implicit LES approaches, in which the discretization error is designed to provide the SFS contribution, e.g. \cite{Grinstein_2007, Moura_2017, Dairay_2017, Fehn_2022, Volpiani_2024}. In this light it is natural to ask how the explicit and implicit SFS contributions in \textcolor{black}{current} SPH methods interact. Is it possible to reduce the significant implicit SFS in \textcolor{black}{current} SPH methods, emerging from statistical physics principles \citep{Posch_1995, Ellero_2010, Borreguero_2019, Okraschevski_2021_1}, using explicit models for $\mathsfbi{\tau}_{SFS}$?} This is the leading theme of this work and in the spirit of similar works in the grid-based LES community \citep{Ghosal_1996,Dairay_2017}.

\section{Methods}
\label{sec:Methods}

There is a large variety of SPH methods today as contrasted by \citet{Eiris_2023}. Hence, care must be taken in the choice of the SPH method for the verification of our incompatibility-hypothesis concerning SPH-LES with eddy viscosity models. We decided to use the locally conservative and second-order accurate meshless finite-mass method (MFM), developed and made publicly available in the open source code GIZMO by \citet{Hopkins_2015}. Since MFM belongs to the large class of \textcolor{black}{current} SPH methods termed as MLS-SPH-ALE \citep{Eiris_2023}, it inherently incorporates both strategies mentioned in Sec. \ref{sec:Motivation} to eliminate the convergence issues of classical SPH. MFM is based on a ALE formulation without particle shifting, which can be operated either in fully Eulerian or quasi-Lagrangian mode, although we will \textcolor{black}{focus on} the latter \footnote{\textcolor{black}{In Appendix \ref{appA} we performed tests on a Cartesian grid in Eulerian mode to investigate whether the incompatibility is generally related to the SPH method or also influenced by the chosen reference frame. Eventually, we realized, by removing the Lagrangian noise and the induced implicit SFS (figure \ref{fig:Rev_Spectrum}), that numerical stability becomes an issue hindering to draw a final conclusion. Yet, we hypothesize that classical eddy viscosity models will withdraw turbulent kinetic energy mostly from scales larger than the kernel even in an Eulerian frame. This would be the natural consequence of the non-locality we criticize.}}. We are convinced that the Lagrangian character is still a key argument for the discretization of fluid flows with \textcolor{black}{current} SPH methods, giving natural access to Lagrangian flow properties like Lagrangian Coherent Structures \citep{Haller_2015, Dauch_2018}. Applying the MFM method to Eqs. (\ref{eq:Mass}), (\ref{eq:Momentum}), (\ref{eq:EOS}) with a second-order accurate MLS approximation of spatial derivatives, one arrives, $\forall i \in \{1,...,N \}$ particles with mass $M_i$, at

\begin{equation}
    \frac{\mathrm{d} M_i} {\mathrm{dt}} = 0 \implies \overline{\rho}_i = M_i \sum_{j=1}^{N_{ngb}}W(\boldsymbol{x}_{i}-\boldsymbol{x}_{j})
    \label{eq:MassParticle}
\end{equation}

\begin{equation}
    M_i \frac{\mathrm{d} \tilde{\boldsymbol{v}}_i} {\mathrm{dt}}  = \sum_{j=1}^{N_{ngb}} - \overline{p}_{ij}^*\boldsymbol{A}_{ij}^{eff} + \left[\tilde{\mathsfbi{\tau}}_{visc,ij}^* + \mathsfbi{\tau}_{SFS,ij}^* \right]\boldsymbol{A}_{ij}^{eff}
    \label{eq:MomentumParticle}
\end{equation}

\begin{equation}
    \overline{p}_i = \overline{p}_{ref} + c_s^2 ( \overline{\rho}_i - \overline{\rho}_{ref} ) ,
    \label{eq:EOSParticle}
\end{equation}
\newline
where we use standard particle notation. Thus, the single index $i$ indicates the numerical proxy of the corresponding field at $\boldsymbol{x}_{i}$. The discretized momentum equation  (\ref{eq:MomentumParticle}) can be interpreted as a Lagrangian finite volume formulation with fluxes to be approximated at effective interface areas $\boldsymbol{A}_{ij}^{eff} \in \mathbb{R}^3$ between particles $i$ and $j$. These interface areas depend on the local particle configuration and the chosen kernel, subsequently the pairing-stable Wendland C4 with $N_{ngb}=128$ as our default \citep{Dehnen_2012}. The interface fluxes, namely $\overline{p}_{ij}^* \in \mathbb{R}$ and $\tilde{\mathsfbi{\tau}}_{visc,ij}^*, \mathsfbi{\tau}_{SFS,ij}^* \in \mathbb{R}^{3 \times 3}$, are computed with approximate Riemann solvers that are slope- and flux-limited \textcolor{black}{and embedded into an explicit single-stage second-order accurate
time integration scheme}. More details on the computation of the effective interface areas $\boldsymbol{A}_{ij}^{eff}$, the Harten-Lax-van Leer contact (HLLC) solver for $\overline{p}_{ij}^*$, the Harten-Lax-van Leer (HLL) solver for $\tilde{\mathsfbi{\tau}}_{visc,ij}^*, \mathsfbi{\tau}_{SFS,ij}^*$ and aspects beyond that can be found in the works of \citet{Hopkins_2015,Hopkins_2016}. We want to note that the resulting system of Eqs. (\ref{eq:MassParticle}), (\ref{eq:MomentumParticle}), (\ref{eq:EOSParticle}) is very similar to the well-known classical weakly-compressible SPH (WCSPH). Thus, we will term it accordingly as weakly-compressible MFM (WCMFM). \textcolor{black}{As a consequence of the explicit weakly-compressible approach, we expect more Lagrangian noise with stronger implicit SFS due to the allowed acoustic waves, comparing with a stable projection-based incompressible approach at the same resolution \citep{Xu_2009,Lind_2012}. This expectation is not SPH agnostic but seems to be a generally anticipated effect as demonstrated by \cite{Vittoz_2019} in a grid-based high-order finite volume context. After all, the resulting WCMFM} discretization is obviously non-local realizing that the differential operators in the momentum balance of equation (\ref{eq:Momentum}) are transferred to $N_{ngb}$ flux evaluations per particle on the kernel scale $D_K$\textcolor{black}{. This} is in the spirit of classical SPH \citep{Du_2020, Vignjevic_2021, Yao_2022}. As will be demonstrated in Sec. \ref{sec:Results}, it seems to be exactly this non-locality 
%in combination with the quasi-Lagrangian particles 
that leads to the incompatibility of SPH-LES with eddy viscosity models for incompressible turbulence \textcolor{black}{(figure \ref{fig:Rev_Spectrum}).}

As a canonical benchmark we will consider a Taylor-Green flow \citep{Taylor_1937} on the periodic domain $\Omega=[0,2\pi]^3$ for three different particle counts of $N\in [128^3, 256^3, 512^3]$. This is a precious test to evaluate the dissipation characteristics of a numerical solver and its ability to resolve incompressible turbulence, e.g. \citep{Brachet_1983, Drikakis_2007,Moura_2017,Dairay_2017,Pereira_2021,Fehn_2022}. As in our former work with classical SPH \citep{Okraschevski_2022}, we will use the DNS solution of \citet{Dairay_2017} at $\Rey=10^4$ as a reference. It was computed with the sixth-order finite difference code Incompact3d \citep{Laizet_2011}. We will initialize and evaluate the WCMFM simulations exactly as we did in \citet{Okraschevski_2022} for classical WCSPH. Thus, we will specify the initial rms Mach number as $Ma_{rms}(t=0\,\text{s})=\sqrt{2e_v(t=0\,\text{s})}/c_s = 0.1$ such that $c_s=5~\text{m}/\text{s}$, $\overline{\rho}_{ref}=1~\text{kg}/{\text{m}^3}$ and $\overline{p}_{ref}=\overline{\rho}_{ref} c_s^2/4 = 6.25~\text{Pa}$ in equation (\ref{eq:EOSParticle}). Consequently, the dynamic viscosity in equation (\ref{eq:StressTensor}) must be $\eta=0.0001~\text{Pas}$ to reach the targeted Reynolds number. We will study and compare the temporal evolution of the averaged kinetic energy $e_v$, its corresponding averaged dissipation rate $\epsilon_t=-\frac{\mathrm{d}e_v}{\mathrm{d}t}$ and the spectral energy density $E(k)$ at $t=14\,\text{s}$. The time instance was selected in accordance with \citet{Dairay_2017}, which ensures that the turbulence is developed and exhibits the expected inertial range scaling of $E(k)\sim k^{-5/3}$ \citep{Kolmogorov_1941, Obukhov_1941, Onsager_1945, Heisenberg_1945} \footnote{Note that there is recent experimental doubt on the quantitative correctness of this scaling in the inertial range \citep{Kuchler_2023}.}. For the evaluation of the spectra, we use the validated methodology of \citet{Bauer_2012} in combination with considerations by \citet{Durran_2017}. This preserves physically important small-scale features of the flow and guarantees the validity of the discrete Parseval relation.

Finally, to work out the interplay between explicit and implicit SFS, we need to specify an explicit SFS model and think about how the implicit SFS contribution can be estimated. For the former, we decided to opt for the $\sigma$-model by \citet{Nicoud_2011}, which is one of the most sophisticated static eddy viscosity models. It eliminates artificial dissipation in two-dimensional flows, laminar shear zones and in case of solid body rotation but likewise shows proper asymptotic scaling near walls \citep{Nicoud_2011, Silvis_2017,Moser_2021}. The explicit model reads in \textcolor{black}{continuous} representation
\begin{equation}
    \mathsfbi{\tau}_{SFS}^{exp}=\eta_{SFS}(\mathsfbi{J}_{\tilde{\boldsymbol{v}}} +\mathsfbi{J}_{\tilde{\boldsymbol{v}}}^{\mathrm{T}}  - \frac{2}{3}\nabla_{\boldsymbol{x}}\cdot \tilde{\boldsymbol{v}}), \quad 
     \eta_{SFS}:=\overline{\rho}(C_\sigma\Delta)^2 \frac{(\sigma_1-\sigma_2)(\sigma_2-\sigma_3)\sigma_3}{\sigma_1^2}~,
     \label{eq:ExplicitSFSTensor}
\end{equation}
with $\sigma_k, k\in\{1,2,3\}$, being the singular values of the tensor $\mathsfbi{J}_{\tilde{\boldsymbol{v}}}^\mathrm{T}\mathsfbi{J}_{\tilde{\boldsymbol{v}}}$, $C_{\sigma}=1.35$ as model constant and the filter width $\Delta=D_K$. The latter is an unambiguous choice emerging from our coarse-graining perspective and a matter of debate in the SPH-LES context \citep{Rennehen_2021, King_2023}. We will elaborate on it more closely in Sec. \ref{sec:Results}. For the estimation of the implicit SFS tensor we assume that density changes of the fluid elements over the kernel scale $D_K$ are much weaker than the corresponding velocity changes. This is reasonable for a weakly-compressible flow developing characteristics of incompressible turbulence and implies that $\tilde{\boldsymbol{v}}=\overline{\boldsymbol{v}}$. Then, with the spatial coarse-graining operation in equation (\ref{eq:Average}) and linearization of the continuum velocity field at a position $\boldsymbol{z}\in \mathbb{R}^3$, one obtains $\tilde{\boldsymbol{v}}(\boldsymbol{x}=\boldsymbol{z},t) = \boldsymbol{v}(\boldsymbol{y}=\boldsymbol{z},t) + \textit{O}(D_K^2)$, which gives a consistency preserving (with respect to the MFM discretization) second-order proxy for the peculiar velocity in equation (\ref{eq:PeculiarVelocity}), namely
\begin{equation}
    \boldsymbol{w}(\boldsymbol{x}_1, \boldsymbol{x}_2, t) \approx \tilde{\boldsymbol{v}}(\boldsymbol{x}_2, t) - \tilde{\boldsymbol{v}}(\boldsymbol{x}_1, t)
    \label{eq:PeculiarVelocityProxy}
\end{equation}
with two different coarse-grained coordinates $\boldsymbol{x}_1,\boldsymbol{x}_2$. Inserting equation (\ref{eq:PeculiarVelocityProxy}) into the SFS tensor in equation (\ref{eq:SubfilterStress}) gives, after discretization of the integral into finite-mass elements, the following estimator for the implicit SFS in particle notation
\begin{equation}
    \mathsfbi{\tau}_{SFS,i}^{imp} \approx - \sum_{j=1}^{N_{ngb}} (\tilde{\boldsymbol{v}}_j-\tilde{\boldsymbol{v}}_i)(\tilde{\boldsymbol{v}}_j-\tilde{\boldsymbol{v}}_i)^\mathrm{T}W(\boldsymbol{x}_i-\boldsymbol{x}_j)M_j~.
    \label{eq:EstimatorImplicitStress}
\end{equation}
Since incompressible turbulence is a convection driven phenomenon, we will evaluate the local importance of the implicit SFS on a particle $i$ in comparison to the coarse-grained convective stress tensor $\overline{\rho}_i\tilde{\boldsymbol{v}}_i\tilde{\boldsymbol{v}}_i^\mathrm{T}$. Therefore, we define the R-index as
\begin{equation}
    R_i:=\frac{|| \mathsfbi{\tau}_{SFS,i}^{imp} ||_F}{||\mathsfbi{\tau}_{SFS,i}^{imp}||_F + ||\overline{\rho}_i\tilde{\boldsymbol{v}}_i\tilde{\boldsymbol{v}}_i^\mathrm{T}||_F}
    \label{eq:R-Index}
\end{equation}
with $||\cdot||_F$ denoting the Frobenius norm.

\section{Results \& Discussion}
\label{sec:Results}

In this section, we will present and critically discuss the results of our study. We start with a qualitative verification of our implementation of the $\sigma$-model \citep{Nicoud_2011} in equation (\ref{eq:ExplicitSFSTensor}) into the code GIZMO \citep{Hopkins_2015}. Therefore, the uniquely colored individual particle IDs are utilized as flow tracers and visualized for the purpose of structure identification (figure \ref{fig:03_Eddy}). Two snapshots before and after the well-known dissipation peak \citep{Brachet_1983}, namely at $t_{1,2}=9\,\pm 3\,\text{s}$, are shown in the first and second column of figure \ref{fig:03_Eddy} for the highest resolution runs $N=512^3$. They vividly render the development of primary instabilities and turbulence in the flow field. The first row displays the case without explicit SFS model (WCMFM), the second with explicit SFS model (WCMFM + SIGMA). In the third row, the central quantity of the $\sigma$-model, namely the eddy viscosity field according to equation (\ref{eq:ExplicitSFSTensor}), is shown. It is scaled by the dynamic viscosity of the flow. The ratio is denoted as $\eta^*$, visually correlates with the flow structures and is evidently non-negligible in the shear flow planes (figure  \ref{fig:03_Eddy}a) where incompressible turbulence develops. In this region the eddy viscosity is dominant over the dynamic viscosity by up to an order of magnitude. This is an anticipated consequence as the coarse-grained viscous stress tensor is bounded from above by the Cauchy-Schwarz inequality and it can be proven that the bound scales with $1/D_K$ \citep{Eyink_2018}. Hence, for regions of underresolved turbulence, in which the kernel scale $D_K$ is larger than viscous length scale, one would expect $\eta^* \gg 1$. A comparison of the flow structures predicted by WCMFM and WCMFM + SIGMA reveals that the explicit SFS model does not inhibit the dynamics of the primary instabilities before the dissipation peak (figure \ref{fig:03_Eddy}a vs. \ref{fig:03_Eddy}c). However, it apparently damps noisy small-scale features in the turbulent flow field after the dissipation peak (figure \ref{fig:03_Eddy}b vs. \ref{fig:03_Eddy}d). This could be incorrectly interpreted as removal of numerically dissipative \textcolor{black}{Lagrangian noise (artificial thermalization)}. As extensively discussed by \citet{Dairay_2017} for grid-based methods, the removal of such an artificial thermalization should be the ultimate goal of an explicit SFS model, eventually resulting in a quantitative improvement for the dissipation characteristics in physical and spectral space. We will show now that the explicit SFS model completely fails with respect to such an quantitative analysis although visually it seems to perform well.
\begin{figure}
\centering
\includegraphics[trim=0cm 0cm 0cm 0cm, clip, width=5in]{./Media/03_EddyViscosity_New.pdf}
\caption{Qualitative verification of the implementation of the $\sigma$-model \citep{Nicoud_2011} for $N=512^3$. (a,b) Flow structures before and after the dissipation peak for the case without explicit SFS model (WCMFM) and (c,d) for the case with explicit SFS model (WCMFM + SIGMA). In (e,f) the scaled eddy viscosity field is displayed.}
\label{fig:03_Eddy}
\end{figure}

\begin{figure}
\centering
\includegraphics[trim=0cm 0cm 0cm 0cm, clip, width=5in]{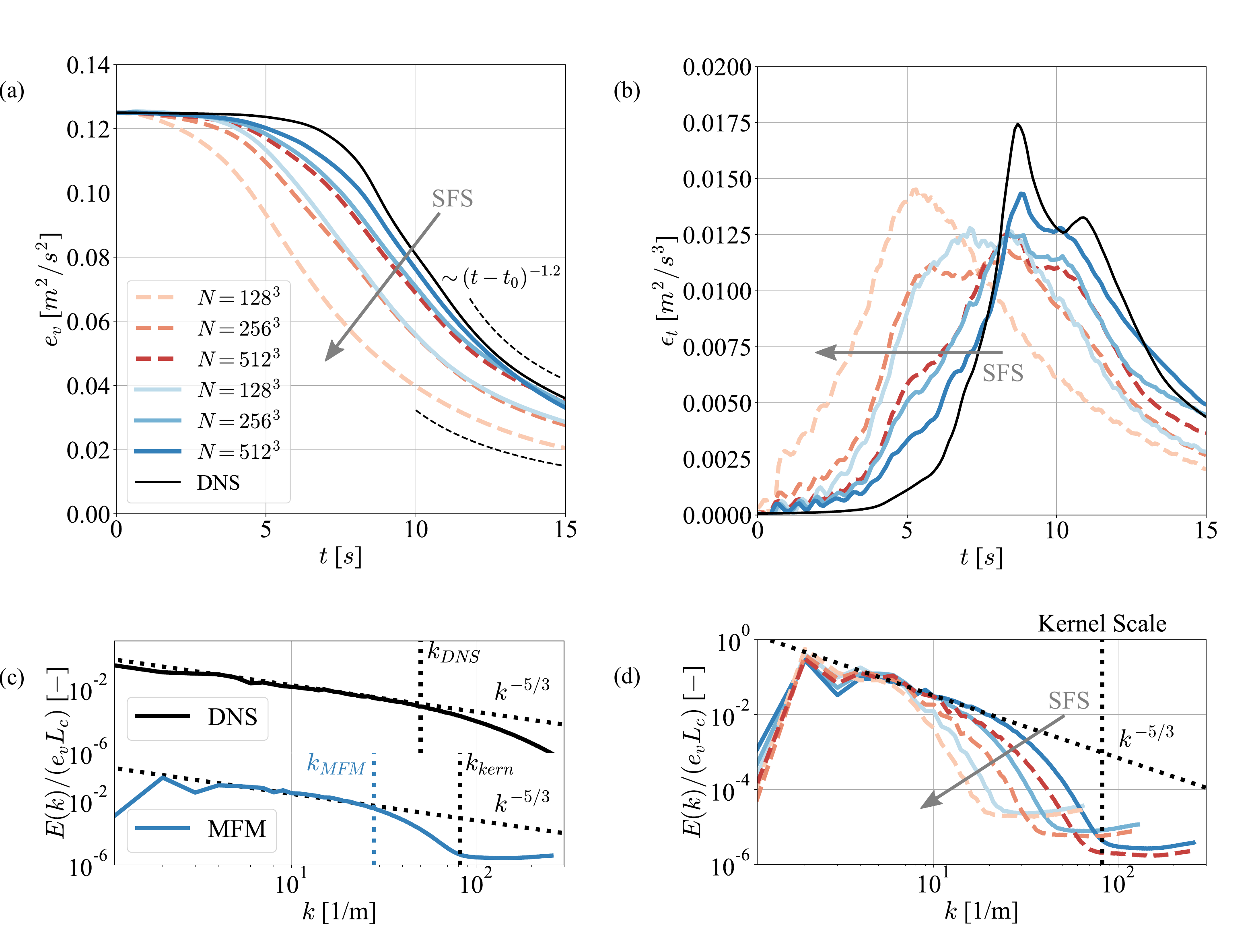}
\caption{Quantitative effect of the $\sigma$-model \citep{Nicoud_2011} in physical and spectral space for different resolutions. (a) Averaged kinetic energy, (b) Averaged dissipation rate, (c) Scaled spectral energy density at $t=14\,\text{s}$ for DNS and WCMFM run ($N=512^3$) without explicit SFS model. (d) Scaled spectral energy density at $t=14\,\text{s}$. For orientation the kernel scale for $N=512^3$ is included.}
\label{fig:01_Convergence}
\end{figure}

Therefore, we compare the temporal evolution of the averaged kinetic energy $e_v$, the averaged dissipation rate $\epsilon_t$ and the spectral energy density $E(k,t= 14\,\text{s})$ in figure \ref{fig:01_Convergence}. The latter is scaled with the corresponding kinetic energy value $e_v(t= 14\,\text{s})$ and $L_c=1\,\text{m}$, such that integration over the wavenumber shells always results in unity. Cases without explicit SFS model (WCMFM) are displayed as solid blue lines, whereas cases with explicit SFS model (WCMFM + SIGMA) are displayed as red dashed lines. It is evident for the kinetic energy evolution in figure \ref{fig:01_Convergence}a that qualitative convergence towards the DNS (solid black line) for increasing particle count can be obtained. After the dissipation peak, as soon as incompressible turbulence develops, the theoretically predicted Saffman decay rate \citep{Skrbek_2000} of $e_v\sim (t-t_0)^{-1.2}$ can be matched (dashed black line). Here, $t_0$ denotes a time shift parameter which accounts for the earlier transitions at lower resolution. These observations are independent of the explicit SFS model and also reflected by the dissipation rate profiles in figure \ref{fig:01_Convergence}b. However, the results clearly demonstrate the detrimental effect of the eddy viscosity model on the dissipation characteristics in physical space. Coincidentally, for the chosen configuration, it seems that the explicit SFS model leads to a fallback of the accuracy by roughly a whole resolution step. The WCMFM cases for $N=128^3$ and $N=256^3$ behave very similar to the WCMFM + SIGMA cases for $N=256^3$ and $N=512^3$. In other words, under ideal code scaling with CFL restriction, to achieve the same result with an explicit SFS model at least an $2^4=16$ times higher computational effort is required. This is a quite drastic finding. Note that for a quasi-Lagrangian particle method the highest resolution WCMFM run ($N=512^3$) without explicit SFS model leads to a comparably accurate prediction of the dissipation characteristics in physical space. Especially, the sharper prediction of the global dissipation peak and the local peak nearby in figure \ref{fig:01_Convergence}b compared to WCSPH with twice as many neighbors \citep{Okraschevski_2022} is prominent. 

Before we proceed with the comparison on the spectral dissipation characteristics for all resolutions, we will first elaborate on the highest resolution WCMFM run in figure \ref{fig:01_Convergence}c. Provably, WCMFM as \textcolor{black}{current} SPH method is able to predict the inertial range scaling of incompressible turbulence by about an order of magnitude in wave number space. However, in comparison to the DNS case the inertial range already terminates at $k_{MFM}<k_{DNS}$ \footnote{By $k_{MFM}$ we denote the effective wavenumber up to which a qualitatively correct spectral behavior can be observed and not the wavenumber corresponding to the mean particle diameter $\Delta l$.}, then passes into the energy deficit range $k\in[k_{MFM};\,k_{kern}]$ and is followed by the \textcolor{black}{Lagrangian noise range} (artificial thermalization) resulting from kernel scale errors. Qualitatively, this is similar to classical SPH and happens although the kernel wavelength satisfies $k_{kern}>k_{DNS}$ \citep{Okraschevski_2022}. It is indicative for the non-local character of the method, the emerging implicit SFS according to equation \ref{eq:EstimatorImplicitStress} and a reaction to peculiar velocities on the kernel scale, which manifest  as artificial thermalization. In the next paragraph this will be detailed, but prior to that we want to show in figure \ref{fig:01_Convergence}d that the explicit SFS model also deteriorates the situation in spectral space due to the non-locality of the method. Instead of removing a significant part of kinetic energy from the artificial thermalization, it dominantly withdraws kinetic energy from the energy deficit range and the partially resolved inertial range. Hence, it erroneously attacks scales which are already badly resolved and not the numerically dissipative ones. While these observations should be sufficient to advise against the usage of classical eddy viscosity models in the SPH-LES context, we want to unravel the interplay of the implicit SFS and the explicit SFS model in the following paragraph and underpin these spectral observations in physical space.

\begin{figure}
\centering
\includegraphics[trim=0cm 2cm 0cm 0cm, clip, width=5in]{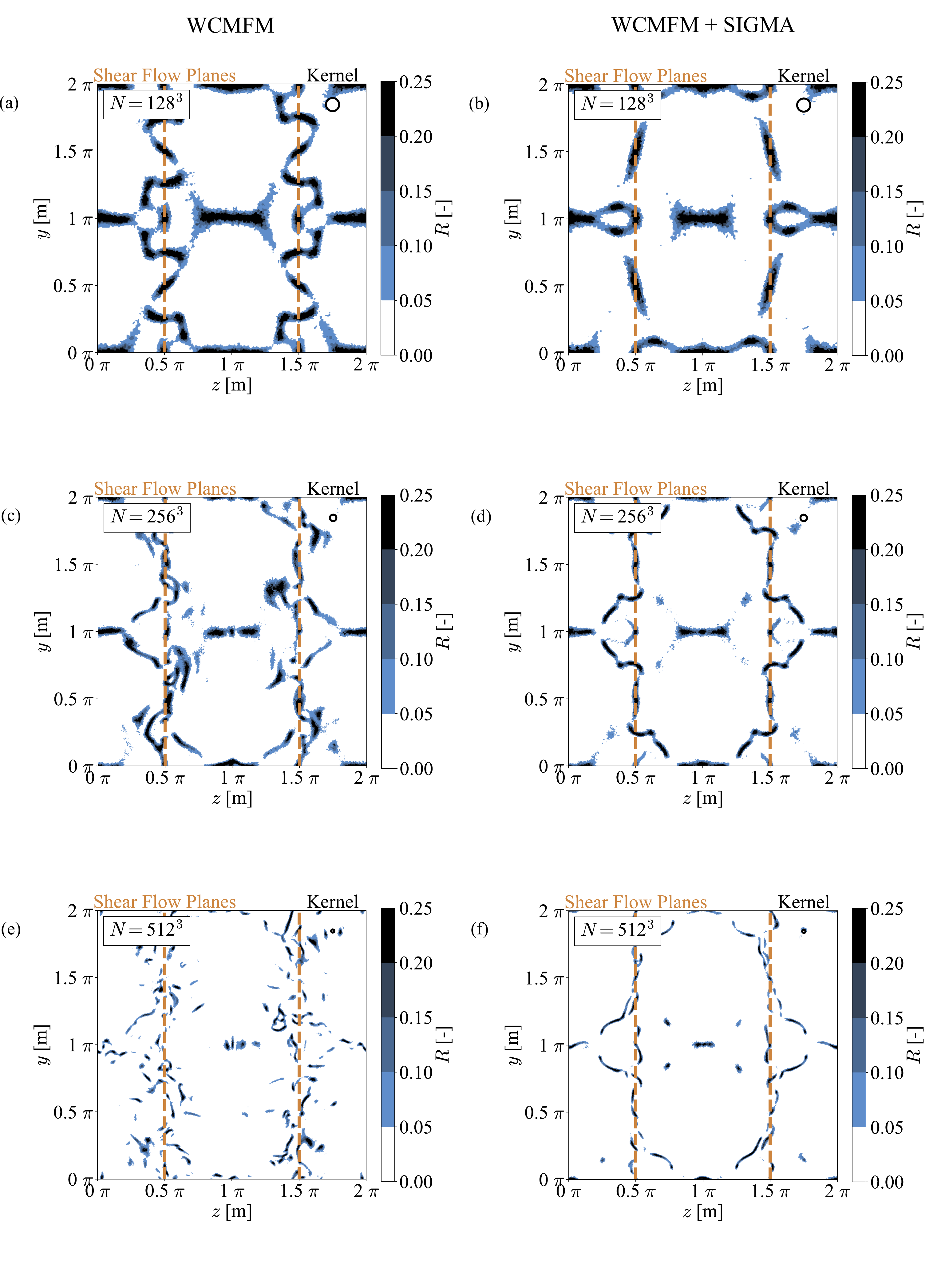}
\caption{Implicit SFS measured by the R-index in equation \ref{eq:R-Index} at the plane $x=\pi$ for the time $t=14\,\text{s}$. Different resolutions are shown without explicit SFS model (WCMFM) and with explicit SFS model (WCMFM + SIGMA).}
\label{fig:05_SFS}
\end{figure}

Ergo, we will visualize the implicit SFS relative to the convective stress using the R-index defined in equation \ref{eq:R-Index} and investigate how the field is affected by resolution and the explicit SFS model. We will focus on the plane $x=\pi$ (figure \ref{fig:03_Eddy}a) for the time $t=14\,\text{s}$, exactly corresponding to the spectra in figure \ref{fig:01_Convergence}c and \ref{fig:01_Convergence}d. The resulting fields are depicted in figure \ref{fig:05_SFS} and contain a kernel element at the given resolution in the upper right corner to assess the extent of emerging coherent structures.
We will start with influence of the spatial resolution for WCMFM without explicit SFS model shown in the first column. Note that due to the definition of the R-index the exact ratio of implicit SFS to the convective stress is given by $|| \mathsfbi{\tau}_{SFS,i}^{imp} ||_F/||\overline{\rho}_i\tilde{\boldsymbol{v}}_i\tilde{\boldsymbol{v}}_i^\mathrm{T}||_F=R_i/(1-R_i)$ and the upper limit of the colorbar indicates that the implicit SFS is a third of the convective stress. Interestingly, for all resolutions, the implicit SFS is only relevant around the shear flow planes where incompressible turbulence develops and forms a coherent network that surpasses the kernel scale. Although, the coherent network becomes more delicate for higher resolution (cp. figure \ref{fig:05_SFS}a and \ref{fig:05_SFS}e), structures with significant contribution remain larger in extent than the given kernel element. Considering that the implicit SFS is a consequence of the peculiar velocities evaluated at the kernel scale but unfolds its effect well beyond the kernel scale, we believe to see the deterministic reason for how the artificial thermalization causes the energy deficit range in spectral space (figure \ref{fig:01_Convergence}c and \ref{fig:01_Convergence}d). Hence, the $|| \mathsfbi{\tau}_{SFS,i}^{imp} ||_F$ field unravels how kernel scale effects propagate to larger scales due to the non-local character of the method. Focusing now on the second column, in which the WCMFM + SIGMA cases are shown, a reflection of the spectral behavior in figure \ref{fig:01_Convergence}d can be clearly seen as well. Even though we would expect for a working  explicit SFS model to significantly reduce $|| \mathsfbi{\tau}_{SFS,i}^{imp} ||_F$ resp. $R$ on the kernel scale, we see a strong non-local damping. The network itself is destroyed, instead of diminishing its amplitude. Recalling the similarity of the  WCMFM cases for $N=128^3$ and $N=256^3$ and WCMFM + SIGMA cases for $N=256^3$ and $N=512^3$ in physical space highlighted above, our requirements on a working explicit SFS model can be specified more precisely. It manifests in the comparison of figure \ref{fig:05_SFS}a vs. \ref{fig:05_SFS}d and figure \ref{fig:05_SFS}c vs. \ref{fig:05_SFS}f. There, the network structure is mainly preserved and the amplitude of $R$ diminished. Together with the spectral statistics in figure \ref{fig:01_Convergence}d, it is evident that this observation correlates with a diminished artificial thermalization and gain in the energy deficit range. This is what we would expect from an explicit SFS model in the SPH-LES context in a much stronger form for a \emph{fixed} resolution but unfortunately see that even sophisticated eddy viscosity models, like the $\sigma$-model by \citet{Nicoud_2011}, fail. 

Taking this new evidence for a \textcolor{black}{current} SPH method into account with our former, congruent results for classical SPH \citep{Okraschevski_2022}, we are convinced that SPH-LES with classical eddy viscosity models is highly detrimental for the prediction of incompressible turbulence. In our opinion this is due to the mismatch of discretization characteristics, namely quasi-Lagrangian particles and non-locality, with the explicit SFS model \textcolor{black}{therefore being spectrally introduced in the already problematic energy deficit range.} 
%assumptions being solely physics motivated. 
We observe the well-known issue from grid-based LES that an \emph{a priori} correct model can perform badly in simulations \emph{a posteriori} \citep{Park_2004,Dairay_2017}. From our coarse-graining perspective \textcolor{black}{current} SPH methods operate intrinsically as  Lagrangian Large Eddy Simulations with implicit SFS for incompressible turbulence. We want to stress that the chosen terminology is not indicative on the quality of this implicit LES approach, which manifests in the spectral energy density in figure \ref{fig:01_Convergence}c. Obviously, it is plagued by an energy deficit range due to artificial thermalization. We have shown already in our former work that the inertial range can be reproduced with a grid-based finite volume Smagorinsky LES at much lower resolution of $N=384^3$ and lower computational cost \citep{Okraschevski_2022}.

\begin{figure}
\centering
\includegraphics[trim=0cm 0cm 0cm 0cm, clip, width=5in]{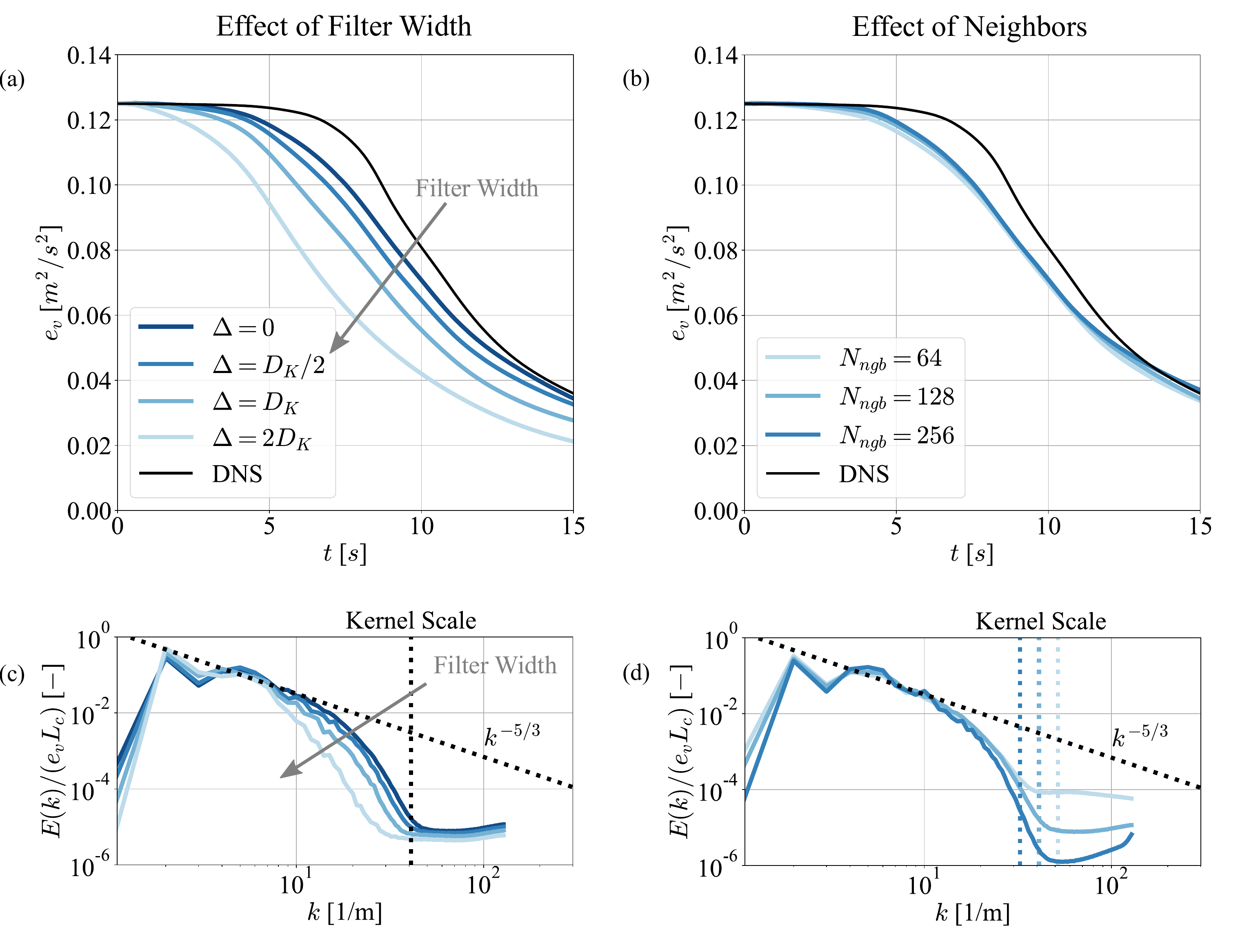}
\caption{(a,c) Influence of the filter width $\Delta$ for WCMFM + SIGMA and (b,d) the neighbor particles $N_{ngb}$ for WCMFM without explicit SFS model. All simulations were performed with $N=256^3$.}
\label{fig:00_LES}
\end{figure}

Before we finally conclude our work, we want to address \textcolor{black}{two more} aspects which are not less important. \textcolor{black}{First, the effect of the filter width $\Delta$ when using an explicit SFS model and, second, the effect of the neighbor particles $N_{ngb}$ for simulations without explicit SFS model. The results are displayed in figure \ref{fig:00_LES} all cases considering $N=256^3$.}

The first aspect is an ongoing matter of debate in the SPH-LES context \citep{Rennehen_2021, King_2023}. Although from our coarse-graining perspective the choice $\Delta=D_K$ is unambiguous, we have considered the two further cases with $\Delta=D_K/2$ \citep{Rennehen_2021} and $\Delta=2D_K$. The case $\Delta=0$ corresponds to the run without an explicit SFS model. Obviously from figure \ref{fig:00_LES}a and figure \ref{fig:00_LES}c, we observe a monotonous trend with increasing filter width. The effect is detrimental in physical space as well as in spectral space, but starts to get irrelevant for $\Delta\leq D_K/2$. This is congruent with the observation of \cite{Rennehen_2021}, who studied forced subsonic turbulence with MFM and different explicit SFS models using a dynamic procedure. It prompts the eventuality that SPH-LES studies with $\Delta$ equal or smaller than the kernel radius, e.g. \cite{Antuono_2021_1,Colagrossi_2021,Lai_2022,King_2023}, just run (slightly) more expensive simulations in which the influence of the explicit SFS model is negligible. 

The second aspect concerns the influence of the number of neighbors $N_{ngb}$ for WCMFM without explicit SFS model. In classical SPH better convergence towards the DNS can be obtained by increasing $N_{ngb}$ \citep{Zhu_2015,Okraschevski_2022}, which is in the spirit of explicitly filtered LES \citep{Lund_2003, Bose_2010}. However, for \textcolor{black}{current} SPH methods as MFM one would expect minor influence of $N_{ngb}$ by construction \citep{Vila_1999,Hopkins_2015}. Indeed, a negligible influence is found according to \ref{fig:00_LES}b and figure \ref{fig:00_LES}d for the well-resolved scales. Merely in spectral space we see a loss of kinetic energy from the artificial thermalization and the energy deficit range which seems irrelevant for the physical space dynamics. This is contrary to classical SPH, practically eliminates $N_{ngb}$ as crucial calibration parameter and is beneficial for the computational cost at finite resolution. Coincidentally, we observe for $N_{ngb}\neq 128$ that the associated kernel scale in figure \ref{fig:00_LES}d is not perfectly separating the energy deficit range and the artificial thermalization as in the former cases.

\section{Concluding Remarks}
\label{sec:Conclusion}

In this study, we presented evidence for the incompatibility of \textcolor{black}{current} SPH methods and classical eddy viscosity models for scale-resolved incompressible turbulence. \textcolor{black}{This result is of particular importance for \textcolor{black}{current} SPH methods, which are often advantageously applied in complex multiphase flows potentially encompassing incompressible turbulence \citep{Colagrossi_2021,Lai_2022,King_2023,Meringolo_2023}.} With our coarse-graining perspective, we could argue and show for MFM, as a representative from the class of MLS-SPH-ALE approaches \citep{Eiris_2023}, that it intrinsically operates as Lagrangian Large Eddy Simulation with significant implicit SFS. Even a sophisticated eddy viscosity model like the $\sigma$-model by \citet{Nicoud_2011} is not able to reduce this implicit SFS due to the non-locality of the discretization method. For the unambiguous filter width $\Delta=D_k$ the explicit SFS model attacks dominantly scales larger than the kernel which are already underresolved. Hence, the explicit SFS model is either highly detrimental physically or, for choices of $\Delta \leq D_K/2$, likely irrelevant and introduces computational overhead. The latter complies with results of \cite{Rennehen_2021}. To our knowledge such a study of the interplay between implicit SFS and explicit SFS model for a \textcolor{black}{current} SPH method is presented for the first time, revealing the familiar \emph{a priori} vs. \emph{a posteriori} dilemma in grid-based LES, e.g. \cite{Park_2004, Dairay_2017}. \textcolor{black}{It elucidates that SPH-LES approaches, even with \textcolor{black}{current} SPH methods, must focus on the mitigation of the implicit SFS to improve the predictive power of the Lagrangian Large Eddy Simulation. Especially one core application area of SPH, namely multiphase flows, could benefit from it if turbulence matters. This progress can be either directly realized by completely new explicit SFS models, which match the discretization characteristics of \textcolor{black}{current} SPH methods, or indirectly by numerical noise mitigation techniques.}

\textcolor{black}{\textcolor{black}{Hypothetically, there exist two in this work disregarded mainstream aspects of current SPH methods compliant with the latter strategy.} Particle shifting \citep{Xu_2009, Lind_2012} and density diffusion \citep{Antuono_2010,Marrone_2011} are established numerical noise mitigation techniques, also in the SPH-LES context \citep{Antuono_2021_2, Antuono_2021_1, Colagrossi_2021, Meringolo_2023}. It is likely that both reduce the peculiar velocities on the kernel scale according to equation \ref{eq:PeculiarVelocityProxy} and indirectly have positive feedback on the implicit SFS according to equation \ref{eq:EstimatorImplicitStress}. However, we believe that the density diffusion, that can be heuristically rationalized by a coarse-graining without the density-weighted Favre velocity \citep{DiMascio_2017}, will be also plagued by non-local effects of the discretization scheme. Very likely, it will not only mitigate noisy but also physically underresolved scales. Nevertheless, this needs to be tested in follow-up studies.}

\textcolor{black}{Another concluding point that needs to be stressed} is that our coarse-graining theory is only valid for $D_K\approx const.$ and $\rho\approx const.$, hence, when compressibility effects are negligible. This is what we ensured by our initial rms Mach number choice $Ma_{rms}(t=0\,\text{s}) = \sqrt{2e_v(t=0\,\text{s})}/c_s = 0.1<0.3$ \citep{Jakobsen_2014} for the decaying Taylor-Green flow at $\Rey=10^4$. Considering this aspect, an essential difference between our results and the results of \citet{Rennehen_2021} should be highlighted. The energy deficit range observed in this study and our former one for classical SPH \citep{Okraschevski_2022}, is there replaced by an energy pile-up known from highly accurate discontinuous Galerkin methods \citep{Moura_2017,Fehn_2022}. We believe this difference is rooted in the Mach number at which the spectral statistics are computed. Whereas $Ma_{rms}\approx0.3$ is controlled by forcing in \citet{Rennehen_2021}, the Mach number in our decaying case can be estimated to be $Ma_{rms}(t=14\,\text{s}) = \sqrt{2e_v(t=14\,\text{s})}/c_s\approx 0.06$. It was already shown by \citet{Hopkins_2015} that the accuracy of the MFM method deteriorates with lower Mach number, possibly explaining the switch from the energy pile-up to the energy deficit also known from classical SPH. Such low Mach numbers are not unusual in engineering applications and favorably comply with our coarse-graining theory.

\textcolor{black}{For completeness we would like to bring up that alternative Lagrangian Large Eddy Simulation approaches exist, which are much better suited to accurately predict incompressible turbulence, if the latter is the exclusive goal. These are the so-called Vortex Particle Methods (VPM) \citep{Alvarez_2024}, which solve the Navier-Stokes equations in their velocity-vorticity form by a decomposition of the fields into self-adaptive Lagrangian vortex elements. Interestingly, these methods \textcolor{black}{suffer} from the exact opposite picture developed in this work for \textcolor{black}{current} SPH methods, namely insufficient implicit SFS to obtain stability. In order to restore the latter, explicit SFS models seem inevitable \citep{Mansfield_1998, Mansfield_1999,Alvarez_2024}. This said, one might wonder whether there is chance to develop a Lagrangian Large Eddy Simulation method to retain the high accuracy of the VPM for incompressible turbulence and flexibility and robustness of \textcolor{black}{current} SPH \textcolor{black}{methods} required for complex multiphase flows. A promising machine-learning based route towards such an ultimate goal could be the one presented by \cite{Woodward_2023} and \cite{Tian_2023}.  }

\backsection[Acknowledgements]{The authors acknowledge support by the state of Baden-Württemberg through bwHPC. Moreover, the authors would like to thank Shreyas Joshi for critical feedback on the manuscript as well as for valuable discussions.}

\backsection[Funding]{This research received no specific grant from any funding agency, commercial or not-for-profit sectors.}

\backsection[Declaration of interests]{The authors report no conflict of interest.}

\backsection[Author ORCIDs]{M. Okraschevski, https://orcid.org/0000-0001-8296-7327; N. Bürkle, https://orcid.org/0000-0002-9380-8716; M. Wicker, https://orcid.org/0009-0009-5255-6936.}

\newpage

\appendix

\section{Incompatibility in the Eulerian reference frame?}\label{appA}

\textcolor{black}{In order to elaborate on the influence of the chosen reference frame, we repeated the numerical experiments presented in Sec. \ref{sec:Results} in a fully Eulerian mode for $N=256^3$. The GIZMO code permits such a course of action as it is methodologically based on a MLS-SPH-ALE framework. The underlying idea is to remove the Lagrangian noise \citep{Hopkins_2015, Lind_2016} and the corresponding implicit SFS by construction to clearly isolate the effect of the explicit SFS model. This allows us \textcolor{black}{to} analyze whether the incompatibility of classical eddy viscosity models with \textcolor{black}{current} SPH methods, rooted in the non-local discretization, persists in an Eulerian reference frame.}

\begin{figure}
\centering
\includegraphics[trim=0cm 0cm 0cm 0cm, clip, width=5in]{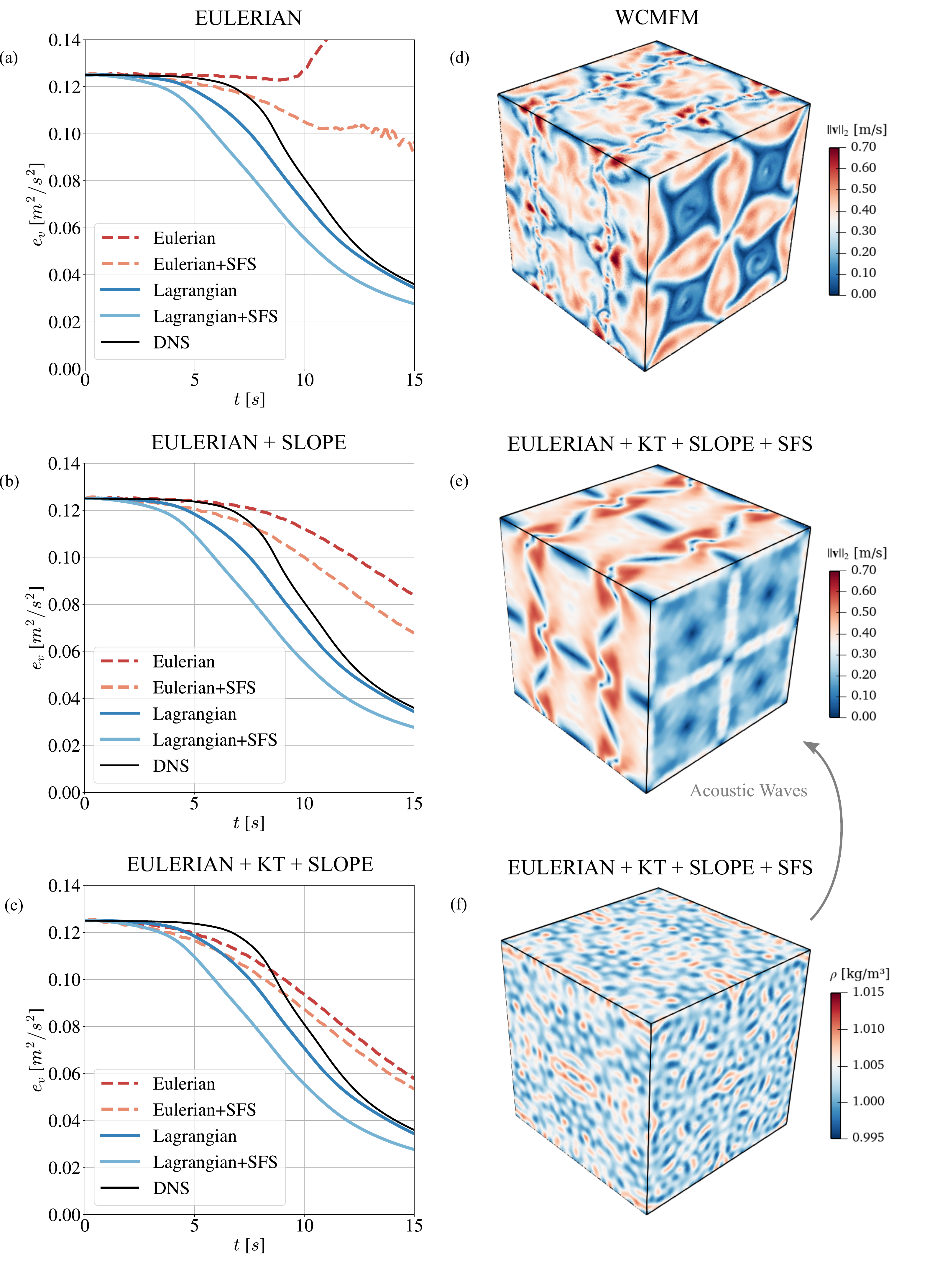}
\caption{Comparison of Lagrangian and Eulerian results for $N=256^3$. (a) Averaged kinetic energy with same parameters. (b) Averaged kinetic energy with conservative slope limiter. (c) Averaged kinetic energy with Kurganov-Tadmor scheme \& conservative slope limiter. (d) Velocity magnitude at $t=14\,\text{s}$ from the Lagrangian WCMFM run without SFS and (e) the Eulerian run with Kurganov-Tadmor scheme \& conservative slope limiter \& SFS. (f) The corresponding density field of the latter. }
\label{fig:Appendix}
\end{figure}

\textcolor{black}{The results for the same parameters and initial conditions are shown in figure \ref{fig:Appendix}a. Obviously, the numerical dissipation solely introduced by the approximate Riemann solvers for the interface fluxes (Sec. \ref{sec:Methods}) is not sufficient to obtain a stable solution, even leading to a blow-up of kinetic energy beyond $t>10\,\text{s}$ in the Eulerian case. Although the $\sigma$-model by \citet{Nicoud_2011} can prevent the blow-up (Eulerian+SFS), it seems not appropriate to eradicate the actual cause of the instability as the kinetic energy starts to plateau beyond $t>10\,\text{s}$, also developing oscillatory behavior. The instability is rooted in constructively interfering acoustic waves leading to density changes of ~30\% deviation from $\overline{\rho}_{ref} = 1~\text{kg}/\text{m}^3$ (not shown herein), which is in contrast to the allowed magnitude of our explicit weakly-compressible approach for the targeted $Ma_{rms}\leq 0.1$ (Sec. \ref{sec:Methods}). We will now show that these interfering acoustic waves cannot be eliminated in the Eulerian reference frame with numerical adaptions in GIZMO, unfortunately hindering to draw a final conclusion in terms of generalization of the incompatibility we observe in the Lagrangian mode. This interference remains even switching to the most dissipative numerical scheme in GIZMO (figure \ref{fig:Appendix}f).}

\textcolor{black}{As noted by \cite{Hopkins_2015}, the slope-limiting procedure specifically designed and calibrated for his MLS-SPH-ALE methods in the Lagrangian reference frame must be likely adapted to more conservative parameters in the context of Cartesian grids in an Eulerian mode. Using the most conservative choice in GIZMO, available as a preprocessor directive (SLOPE\_LIMITER\_TOLERANCE=0), results in the averaged kinetic energy evolution in figure \ref{fig:Appendix}b. Although this gives a stable result, we obtain seemingly sub-dissipative behavior in the kinetic energy after $t>6\,\text{s}$ and adding the explicit SFS model does not lead to qualitative improvement. However, realizing that the Eulerian run without SFS can perfectly match the kinetic energy characteristics up to $t\leq6\,\text{s}$, we additionally exchanged the HLLC flux approximation by the central Kurganov-Tadmor scheme \citep{Kurganov_2000, Panuelos_2020} without the novel dissipation switch developed by \cite{Panuelos_2020}. This should give the most dissipative Eulerian run on a Cartesian grid in GIZMO and hopefully eliminate the seemingly sub-dissipative behavior observed before. \textcolor{black}{Evidently from figure \ref{fig:Appendix}c, the overall dissipation increases from the beginning but the sub-dissipative behavior persists while it is shifted to $t>9\,\text{s}$}. Again, the explicit SFS model does not qualitatively improve this situation. To rationalize these effects, the magnitude of the velocity fields from the least dissipative Lagrangian run (figure \ref{fig:Appendix}d) and the most dissipative Eulerian run (figure \ref{fig:Appendix}e) can be compared at $t=14\,\text{s}$, where incompressible turbulence should be present. Apparently, taken all dissipation measures into account, the Eulerian velocity field is much smoother than the Lagrangian reference, even showing a laminarization of the dynamics in the shear flow planes. This is surprising, given that the averaged kinetic energy levels exceed the Lagrangian ones after $t>9\,\text{s}$ (figure \ref{fig:Appendix}c). Actually, a decrease of the (effective) Reynolds number should result in a monotonous decrease in the averaged kinetic energy levels \citep{Fehn_2022}. Hence, kinetic energy must be introduced by another mechanism. These are the constructively interfering acoustic waves mentioned above, which manifest in the corresponding density field in figure \ref{fig:Appendix}f. Although their magnitude is reasonable in terms of the weakly-compressible approach taken all dissipation measures into account, they are the dominant, unphysical feature in the density field. Unfortunately, they also introduce oscillatory behavior in the velocity field (figure \ref{fig:Appendix}e) and are the root of instability in figure \ref{fig:Appendix}a and the seemingly sub-dissipative behavior in figure \ref{fig:Appendix}b, \ref{fig:Appendix}c.}

\textcolor{black}{From these observations we extract three main conclusions:}

\begin{itemize}
    \item \textcolor{black}{To perform implicit LES of incompressible turbulence with the MLS-SPH-ALE methods of \cite{Hopkins_2015}, the Lagrangian reference frame 
    with its Lagrangian noise and the related implicit SFS seem inevitable. In the Eulerian frame the numerical schemes lack a direct dissipation mechanism for the detrimental acoustic waves.}\textcolor{black}{
    \item If an implicit LES in an Eulerian reference frame would be possible, we would expect that the incompatibility of classical eddy viscosity models and \textcolor{black}{current} SPH methods persists as it is rooted in the non-local discretization.}
    \item \textcolor{black}{The $\sigma$-model by \citet{Nicoud_2011} does not resolve the acoustic wave issue due to a lack of awareness. This could also be interpreted as a sort of incompatibility but is different from the non-local incompatibility we refer to in our work.}
\end{itemize}

\bibliographystyle{jfm}
\bibliography{jfm}

\end{document}